# Brain MR-Elastography in microgravity analogous conditions
Fatiha Andoh[1], Claire Pellot-Barakat[1], and Xavier Maître[1]
[1]Université Paris-Saclay, CEA, CNRS, Inserm, BioMaps, Orsay, France

## Synopsis

The head down tilt (HDT) position is commonly used to simulate vascular and tissue fluid dynamics during spaceflights. In HDT position, the cerebral autoregulation faces difficulties to adjust the vascular tone while the cephalad fluid shifts may yield increased intracranial pressures and altered mechanical properties. Recent MRI T2 mapping in HDT position have shown fluid overpressure in the brain and resulting loss of water contents in the CSF and orbital compartments. Brain MRE was performed here in similar HDT conditions. It was sensitive enough to provide new insights on the overall mechanical response of brain tissues in microgravity analogous conditions.

## Summary of Main Findings/Short Synopsis

Brain fluid overpressure and resulting loss of water contents in CSF and orbital compartments were confirmed by $T_2$ mapping in head down tilt position. The overall brain mechanical response in such microgravity analogous conditions, cerebral tissue stiffening, was revealed by whole brain MRE.

## Introduction

Travelling in space affects the terrestrial human organism. The microgravity environment noticeably reconditions the subtle balance of fluid pressures and flows throughout the body.[1] These conditions are mimicked on the ground by positioning subjects head down. Reductions of brain water content have already been reported in head-down tilt (HDT) position by MRI $T_1$ or $T_2$ mapping.[2,3] Here, brain MRE was performed to probe the mechanical properties of the brain tissue under such controlled gravity-driven pressure variations.

## Methods

One healthy subject (male, 49 y/o) was imaged first in supine (0°) then in HDT (17°) positions after a 40 min rest in each position for proper fluid redistribution. In-between the two MRI acquisitions, the subject was asked to walk for 30 min. The HDT position was established with a tilted board placed at 17° onto the MR bed.

Measurements were performed using a standard head SENSE coil in a 1.5 T Achieva MR system (Philips, Best, Netherlands). First, a multi spin-echo sequence was applied for $T_2$ mapping with *FOV* = (210×210×154) mm$^3$, voxel = (2 mm)$^3$, and *TE/TR* = {20,40,60,80,100}/10,000 ms. Second, remotely generated pressure waves were guided into the buccal cavity to induce shear waves throughout the brain[4] while applying a synchronized motion-sensitized spin-echo sequence to record the resulting displacement fields with *FOV* = (210×210×152) mm$^3$, voxel = (2.94 mm)$^3$, and *TE/TR* = 29/2,000 ms. The excitation mechanical frequency was set at 104 Hz, close to a system resonant mode, to achieve optimal wavelength sampling for MRE reconstruction.[5]

$T_2$ maps were computed from the fitting of the five TE images and registered before inferring the absolute $T_2$ variation maps: $\Delta T_2 = |T_2^{17°} - T_2^{0°}|$ and the relaxivity relative variation $\Delta R_2 = (R_2^{17°} - R_2^{0°})/R_2^{0°}$. Shear velocity maps, $V_s$, were extracted from the displacement fields acquired in the three spatial directions by inversion of the Helmoltz equation of the displacement field rotational, **q**. The shear dynamic and loss moduli, $G'$ and $G''$, were then deduced. The first echo image was segmented using SPM12 (UCL, United-Kingdom) to infer masks of cerebral grey and white matters, cerebellum, and CSF. Relative variations of the median values of $R_2$, $V_s$, $G'$, and $G''$ between the 0° and 17° positions were calculated in every axial slice.

## Results

$T_2$ is essentially the same for both positions in the cerebrum and cerebellum but it decreases at 17° in the CSF compartment ($\Delta T_2 \sim 50$ ms) and probably in the orbital compartment (Figure 1, Table 1).

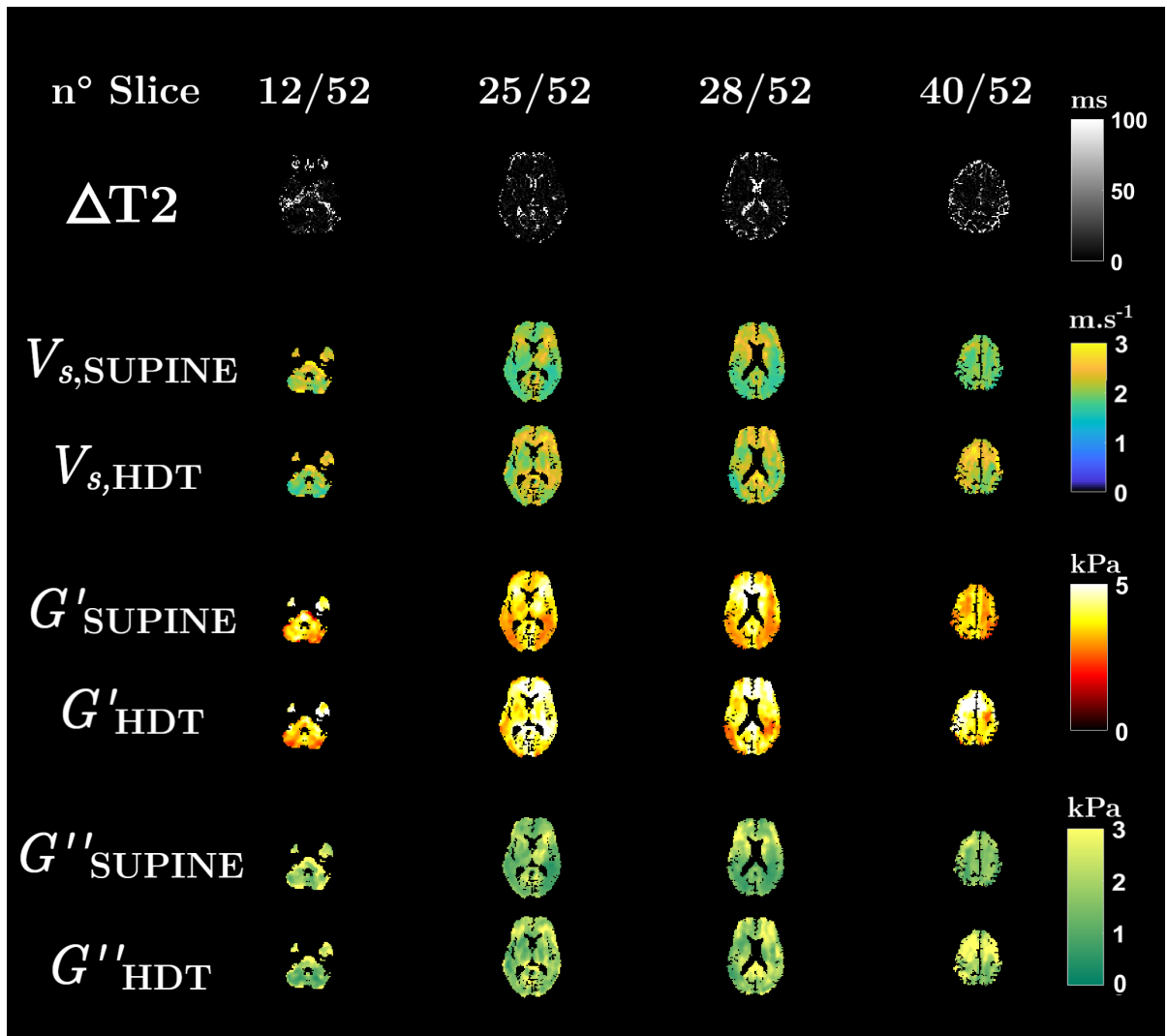

Figure 1: Axial maps at slices 12, 25, 28 and 40 of the absolute variation of the MR signal lifetime, $\Delta T_2$, between 0° and 17° positions (top row). $\Delta T_2$ is essentially zero everywhere but in the CSF and orbital compartments where it exhibits a clear $T_2$ decrease in HDT position. Shear velocity, $V_s$, and viscoelastic

moduli, $G'$ and $G''$, maps reveal a global mechanical increase between 0° supine and 17° HDT in the cerebral tissues (bottom rows).

In Figure 2, the relative variation of the median relaxivity displays clear positive peaks around the fourth ventricle (slices 5-7), the eyes (slices 9-13), the third and the lateral ventricles (slices 25-31), and the subarachnoid CSF (slices 41-43).

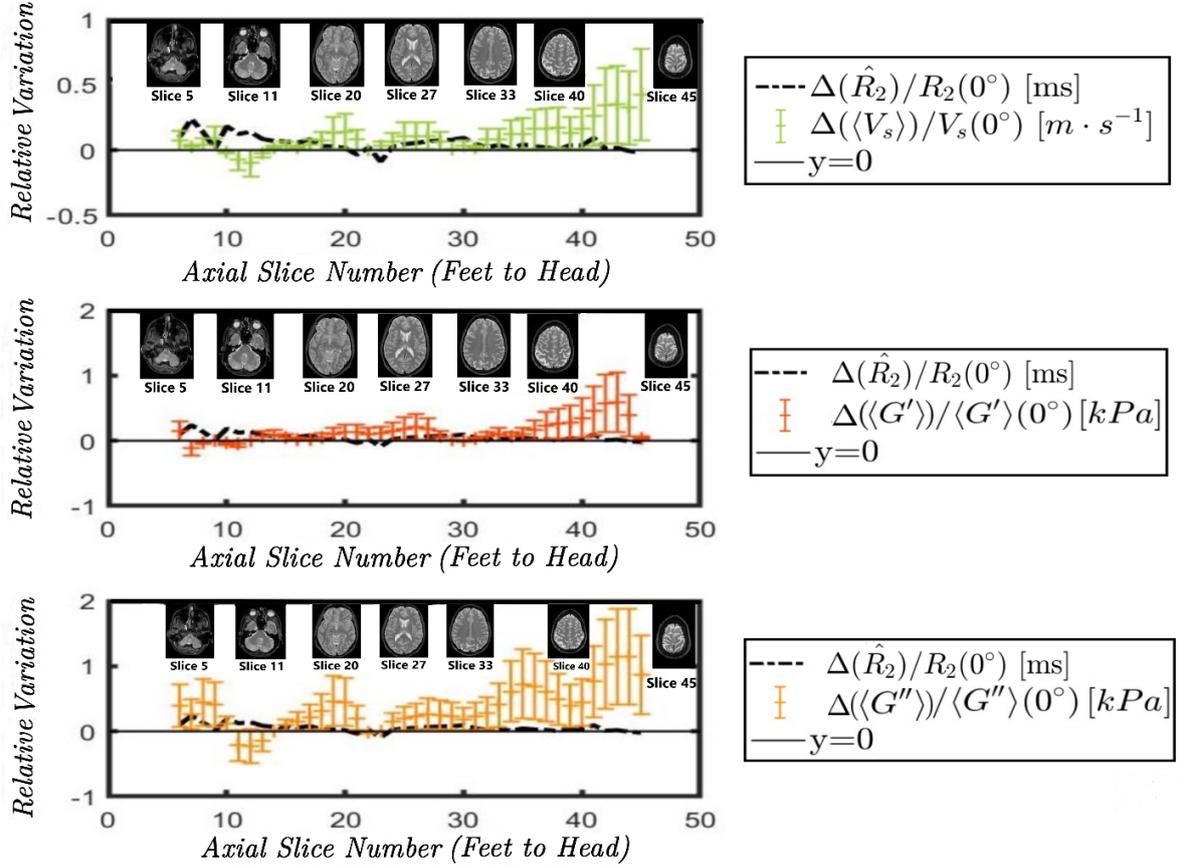

Figure 2: Relative variation of the median MR transverse relaxivity ($R_2 = 1/T_2$) (black), of the mean shear velocity ‹$V_s$› (green), the mean shear elasticity ‹$G'$› (dark orange) and shear viscosity ‹$G''$› (light orange) across the feet-head axial slices. While $\hat{R}_2$ positively varies only locally in the CSF and orbital compartments, ‹$V_s$›, ‹$G'$› and ‹$G''$› increase everywhere in the cerebral tissues, remain rather constant in cerebellum and decrease around the tentorium.

Brain MRE datasets in both positions exhibit similar SNR (~45), displacement field amplitudes ($A_T$~7.7 μm), and **q**-based quality factors ($Q$~46) (Table 1) whereas the extracted shear velocity and viscoelastic moduli globally increase everywhere in the cerebrum (Tables 2 and 3) between 0° and 17° positions.

| Brain | ⟨SNR⟩$_B$ | ⟨$A_T$⟩$_B$ [μm] | ⟨$Q$⟩$_B$ | ⟨$T_2$⟩$_B$ [ms] | ⟨$V_S$⟩$_B$ [m·s$^{-1}$] | ⟨$G'$⟩$_B$ [kPa] | ⟨$G''$⟩$_B$ [kPa] |
|---|---|---|---|---|---|---|---|
| 0° supine | 45 | (7.7±2.9) | (47±32) | 235 | (2.12±0.29) | (3.71±0.67) | (1.81±0.79) |
| 17° HDT | 45 | (7.9±2.8) | (44±31) | 216 | (2.29±0.32) | (4.20±0.83) | (2.27±0.92) |
| Variation | ~ | ~ | ~ | ↓ | ↑ | ↑ | ↑ |

Table 1: Overall brain mean values for SNR, displacement field amplitude, $A_T$, quality factor, $Q$, transverse relaxation time $T_2$, shear velocity, $V_s$, and shear viscoelasticity moduli, $G'$ and $G''$, at 0° and 17°.

Yet they remain the same in the cerebellum (Table 4) and even decrease around the tentorium cerebelli (Figures 1-2). The cerebral mechanical increase ($V_s$, $G'$, and $G''$) is rather scattered throughout voxels but it follows a general positive gradient towards the superior part of the brain (Figure 2).

| Cerebrum | White matter | | | |
|---|---|---|---|---|
| | $\langle T_2 \rangle_{WM}$ [ms] | $\langle V_s \rangle_{WM}$ [m·s$^{-1}$] | $\langle G' \rangle_{WM}$ [kPa] | $\langle G'' \rangle_{WM}$ [kPa] |
| 0° supine | 110 | (2.13±0.27) | (3.79±0.69) | (1.84±0.74) |
| 17° HDT | 112 | (2.31±0.29) | (4.34±0.78) | (2.25±0.86) |
| Variation | ~ | ↑ | ↑ | ↑ |

Table 2: Cerebral white matter mean values for transverse relaxation time $T_2$, shear velocity, $V_s$, and shear viscoelasticity moduli, $G'$ and $G''$, at 0° and 17°.

| Cerebrum | Grey matter | | | |
|---|---|---|---|---|
| | $\langle T_2 \rangle_{GM}$ [ms] | $\langle V_s \rangle_{GM}$ [m·s$^{-1}$] | $\langle G' \rangle_{GM}$ [kPa] | $\langle G'' \rangle_{GM}$ [kPa] |
| 0° supine | 89 | (2.10±0.26) | (3.70±0.63) | (1.72±0.69) |
| 17° HDT | 91 | (2.30±0.29) | (4.22±0.78) | (2.30±0.85) |
| Variation | ~ | ↑ | ↑ | ↑ |

Table 3: Cerebral white and grey matter mean values for transverse relaxation time $T_2$, shear velocity, $V_s$, and shear viscoelasticity moduli, $G'$ and $G''$, at 0° and 17°.

| Cerebellum | $\langle T_2 \rangle_{Cb}$ [ms] | $\langle V_s \rangle_{Cb}$ [m·s$^{-1}$] | $\langle G' \rangle_{Cb}$ [kPa] | $\langle G'' \rangle_{Cb}$ [kPa] |
|---|---|---|---|---|
| 0° supine | 108 | (2.19±0.43) | (3.56±0.74) | (2.11±1.15) |
| 17° HDT | 109 | (2.19±0.44) | (3.55±0.94) | (2.10±1.28) |
| Variation | ~ | ~ | ~ | ~ |

Table 4: Cerebellum mean values for transverse relaxation time $T_2$, shear velocity, $V_s$, and shear viscoelasticity moduli, $G'$ and $G''$, at 0° and 17°.

## Discussion

The increase of relaxivity between 0° and 17° positions corroborates the results found by Caprihan *et al.* at 13°, which were interpreted as a reduction of water content in the eyes and in the subarachnoid CSF. For the authors, it resulted from an exudation of free water with the pressure increase in those compartments relative to the surrounding tissue.[2] It is extended here at 17° to the overall ventricular region where the CSF flows, potentially as a result of a fluid higher pressure. No difference could be measured otherwise in the brain tissue.

On the contrary, the increase of $V_s$, $G'$, and $G''$ in the cerebral grey and white matters reveals the mechanical repercussion of the HDT position, which impacts the whole cerebrum. This effect continuously intensifies moving towards the top of the brain. The overall cerebral mean velocity increase is only 10% but the velocity variation can reach more than 50% in the top slices. This effect is even more pronounced for $G'$ and $G''$. In the cerebellum, away from the CSF overpressure, no effect is observed, and around the tentorium, an opposite effect occurs as if the inversion of the gravity may also relieve the weight of the cerebrum.

## Conclusion

In microgravity analogous conditions, MRI $T_2$ mapping underlines the fluid overpressure in the brain and the resulting loss of water contents in the CSF and orbital compartments only. MRE provides complementary information on the overall brain mechanical response. Cerebral tissue stiffening is underlined by the increase of the mechanical parameters in HDT towards the superior regions of the cerebrum, which may result from both intracranial overpressure and gravity dependence. MRE can be sensitive to capture these changes locally and could be advantageously used to detect other mechanical changes due to pathological phenomena. HDT position could serve as a benchmark for brain MRE.

## Acknowledgements

MRE experiments were performed on the 1.5 T MRI platform of CEA/SHFJ affiliated to the France Life Imaging network (grant ANR-11-INBS-0006).